\documentclass[useAMS]{emulateapj}
\usepackage[]{natbib}
\newcommand\exo{EXO~2030+375}
\usepackage{graphicx}

\def\lsim{\lower.5ex\hbox{$\; \buildrel < \over \sim \;$}}
\def\gsim{\lower.5ex\hbox{$\; \buildrel > \over \sim \;$}}

\shorttitle{Suzaku Observation of EXO~2030+375}
\shortauthors{Naik et al.}

\begin{document}
\title{Timing and Spectral properties of Be/X-ray pulsar EXO~2030+375 during a
type~I outburst}

\author{Sachindra Naik\altaffilmark{1}, Chandreyee Maitra\altaffilmark{2}\nonumber\footnote{Joint Astronomy Programme, Indian Institute of Science, Bangalore - 560012, India}, Gaurava K. Jaisawal\altaffilmark{1}, Biswajit Paul\altaffilmark{2}}

\affil{\altaffilmark{1} Astronomy \& Astrophysics Division, Physical Research Laboratory, Ahmedabad - 380009, India}
\affil{\altaffilmark{2} Raman Research Institute, Sadashivnagar, C. V. Raman Avenue, Bangalore - 560080, India}
\email{snaik@prl.res.in}

\begin{abstract}
We present results from a study of broadband timing and spectral 
properties of EXO~2030+375 using a $Suzaku$ observation. Pulsations with a 
period of 41.41 s and strong energy dependent pulse profiles were clearly 
detected up to ~100 keV. Narrow dips are seen in the profiles up to $\sim$70 keV. 
Presence of prominent dips at several phases in the profiles up to such high 
energy ranges were not seen before. At higher energies, these dips gradually 
disappeared and the profile appeared single-peaked. The 1.0-200.0 keV  broad-band
spectrum is found to be well described by a partial covering high energy cut-off 
power-law model. Several low energy emission lines are also detected in the pulsar 
spectrum. We fitted the spectrum using neutral as well as partially ionized absorbers 
along with above continuum model yielding similar parameter values. The partial 
covering with partially ionized absorber resulted into marginally better fit.
The spectral fitting did not require any cyclotron feature in the best fit model. 
To investigate the changes in spectral parameters at dips, we carried out 
pulse-phase-resolved spectroscopy. During the dips, the value of additional 
column density was estimated to be high compared to other pulse phases. 
While using partially ionized absorber, the value of ionization parameter is also 
higher at the dips. This may be the reason for the presence of dips up to higher 
energies. No other spectral parameters show any systematic variation with pulse 
phases of the pulsar.
\end{abstract}


\section{Introduction}

Be/X-ray binaries represent the largest subclass of high mass X-ray binary
systems. The compact object in these systems is generally a neutron star 
(pulsar) whereas the companion is a B or O-type star which shows Balmer 
emission lines in its spectra. The binary optical companion lies well within 
the Roche lobe. The objects in these binary systems are typically in a 
wide orbit with moderate eccentricity. Though evolutionary model calculations 
show that binary systems with white dwarf and Be star or black hole and Be star 
should also exist, clear evidence of the existence of such binary systems has 
not been found as yet (Zhang, Li \& Wang 2004 and references therein). The 
neutron star in these Be/X-ray binary systems accrets matter while passing 
through the circumstellar disk of the companion Be star. The abrupt accretion 
of matter onto the neutron star while passing through the circumstellar disk 
of the Be companion or during the periastron passage results in strong X-ray 
outbursts (Okazaki \& Negueruela 2001). During such outbursts, the X-ray 
emission from the pulsar can be transiently enhanced by a factor more than 
$\sim$10.  Be/X-ray binary systems generally show periodic normal (type~I) X-ray 
outbursts that coincide with the periastron passage of the neutron star and 
giant (type~II) X-ray outbursts which do not show any clear orbital dependence 
apart from the persistent low luminosity X-ray emission during quiescent (Negueruela 
et al. 1998). The neutron stars in the Be/X-ray binary systems are found to be 
accretion powered X-ray pulsars except a very few cases such as LS~I+61303 (Massi 
et al. 2004). The spin period of these pulsars is found to be in the range of a 
few seconds to several hundred seconds. The X-ray spectra of these pulsars are 
generally hard. Fluorescent iron emission line at 6.4 keV is observed in the 
spectrum of most of the accretion powered X-ray pulsars. For a brief review on 
the properties of Be/X-ray binary pulsars, refer to Paul \& Naik (2011).

The transient X-ray pulsar \exo~ was discovered during a giant outburst in 
1985 with $EXOSAT$ observatory (Parmar et al. 1989a). Optical and near-infrared 
observations of the $EXOSAT$ error circle identified a B0 Ve star as the 
counterpart of \exo~ (Motch \& Janot-Pacheco 1987; Coe et al. 1988). Using the 
$EXOSAT$ observations in 1985, the spin and orbital periods of the pulsar 
were estimated to be 42 s and 44.3-48.6 days, respectively. Analyzing $BATSE$ 
monitoring data of several consecutive outbursts of the pulsar \exo~ in 1992, 
Stollberg et al. (1997) derived following orbital parameters of the binary 
system : orbital period P$_{orb}$ = 46.02$\pm$0.01 days, $e$ = 0.36$\pm$0.02, 
$a_x$sin~$i$ = 261$\pm$14 lt-sec, $\omega$ = 223$^\circ$.5 $\pm$ 1$^\circ$.8, 
and time of periastron passage $\tau$ = 2448936.8 $\pm$ 0.3 days. 
During the giant outburst in 1985, the pulsar was observed with $EXOSAT$ 
observatory. A significant change in 1-20 keV luminosity by a factor of 
$\geq$2500 was detected compared to that during the quiescent phase.
A dramatic change in pulse period was seen during the luminosity decline 
with spin-up timescale of $-P/\dot{P} \sim$ 30 yr (Parmar et al. 1989a). 
During the outburst, the pulse profile of the pulsar was found to be strongly 
luminosity dependent. At high luminosity, the pulse profile consisted of one 
main pulse and a small inter-pulse, separated by $\sim$180$^\circ$ phase. The 
strength of the two pulses was reversed when the luminosity was decreased by 
a factor of $\sim$100 (Parmar et al. 1989b). By using a geometric model, 
Parmar et al. (1989b) explained that the dominant beam of emission changed 
from a fan-beam to a pencil-beam during the decrease in luminosity and that 
resulted in the strength reversal of the main and inter-pulse. An extensive 
monitoring campaign of \exo~ with $BATSE$ and Rossi X-ray Timing Explorer 
($RXTE$) showed that a normal outburst has been detected for nearly every 
periastron passage for $\sim$13.5 years (Wilson, Fabregat \& Coburn 2005). 

The spectral analysis of \exo~ had been carried out by using $EXOSAT$ data 
during outburst (Reynolds, Parmar \& White (1993); Sun et al. (1994)). The  
1-20 keV spectrum was described by a two component continuum model consisting
of a blackbody component with temperature $\sim$1.1 keV whereas the hard
X-ray part was represented by a power-law. $RXTE$ monitoring of the pulsar 
during an outburst in 1996 June-July also suggested that a two component
(blackbody and power-law with an exponential cut-off) model was required 
to describe the 2.7--30 keV pulsar spectrum (Reig \& Coe 1999). A spectral
feature at $\sim$36 keV in the hard X-ray spectrum (in 17-65 keV range) was 
ascribed to a possible cyclotron absorption line implying the estimated 
magnetic field of the pulsar to be 3.1$\times$10$^{12}$ Gauss (Reig \& Coe 1999). 
However, using regular monitoring data of \exo~ with the $RXTE$ from 2006 June 
to 2007 May, covering the first giant outburst since its discovery in 1985, Wilson 
et al. (2008) reported a cyclotron feature at $\sim$11 keV and estimated the 
magnetic field strength to be 1.3$\times$10$^{12}$ Gauss. This feature was
consistently detected in the pulsar spectrum for about 90 days when the 
2-100 keV luminosity was above 5$\times$10$^{37}$ erg s$^{-1}$. $INTEGRAL$
and $Swift$ observation of the same giant outburst was used to describe
the 3-120 keV spectra by using an absorbed power-law with an exponential
cut-off, an iron emission line and some peculiar features in 10-20 keV energy
range. The feature in 10-20 keV energy range was modeled by a broad emission line at 
$\sim$13-15 keV or by two absorption lines at $\sim$10 keV and $\sim$ 20
keV (Klochkov et al. 2007). Pulse-phase resolved spectroscopy of $INTEGRAL$
observation of the pulsar during the giant outburst showed significant spectral
variability of the continuum parameters (Klochkov et al. 2008). In the process,
evidence of the presence of an absorption line at $\sim$63 keV was found at a 
narrow pulse phase interval when \exo~ was at the peak of its giant outburst. 
This feature was interpreted as the harmonic of the previously reported 
$\sim$36 keV cyclotron line. 

For a detailed study of timing and spectral properties, \exo~ was observed
with $Suzaku$ on 14 May 2007, at the peak of a regular Type~I outburst. The results 
obtained from the timing and spectral analysis of the $Suzaku$ observation are 
presented in this paper.

\begin{figure}
\centering
\includegraphics[height=3.4in, width=3in, angle=-90]{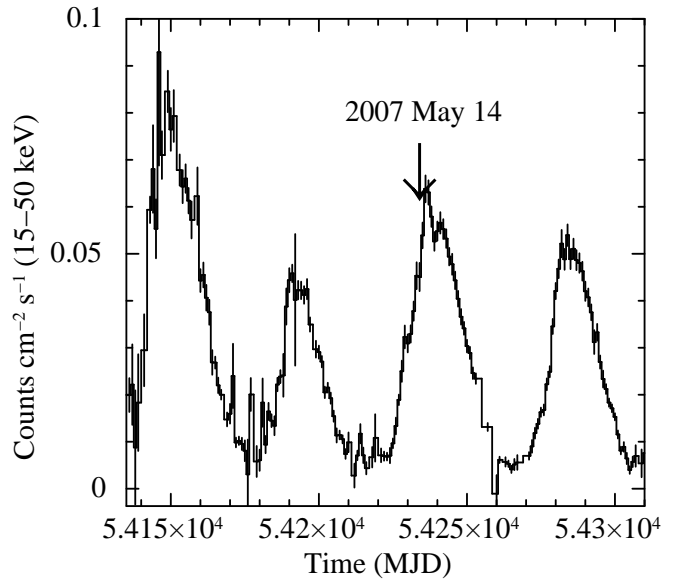}
\caption{The $Swift$/BAT light curve of \exo~ in 15-50 keV energy band,
from 2007 February 4 (MJD 54135) to 2007 July 29 (MJD 54310). The arrow
mark shows the date of the $Suzaku$ observation of the pulsar.}
\label{fg1}
\end{figure}

\section{Observation}

Observation of the transient Be/X-ray binary pulsar \exo~ was carried out
on 2007 May 14 at the peak of its regular type~I outburst. The publicly
available arcival data (ver-2.0.6.13) of above observation was used in the
present work to investigate the properties of the pulsar during the outburst.
Figure~\ref{fg1} shows the one-day averaged light curve of \exo~ in 15-50 keV 
energy range obtained from the $Swift$/BAT monitoring data between 2007 February 4 
and 2007 July 29 covering the present type-I outburst. The arrow mark in the figure 
shows the $Suzaku$ observation of the pulsar during the peak of the outburst. This 
observation was carried out at ``HXD nominal'' pointing position for effective 
exposures of $\sim$57 ks and $\sim$53 ks for XIS and HXD, respectively. The XIS 
were operated with ``burst'' clock mode in ``1/4 window'' option, covering  
17$'$.8$\times$4$'$.4 field of view. 

\begin{figure}
\centering
\includegraphics[height=3.1in, width=4.in, angle=-90]{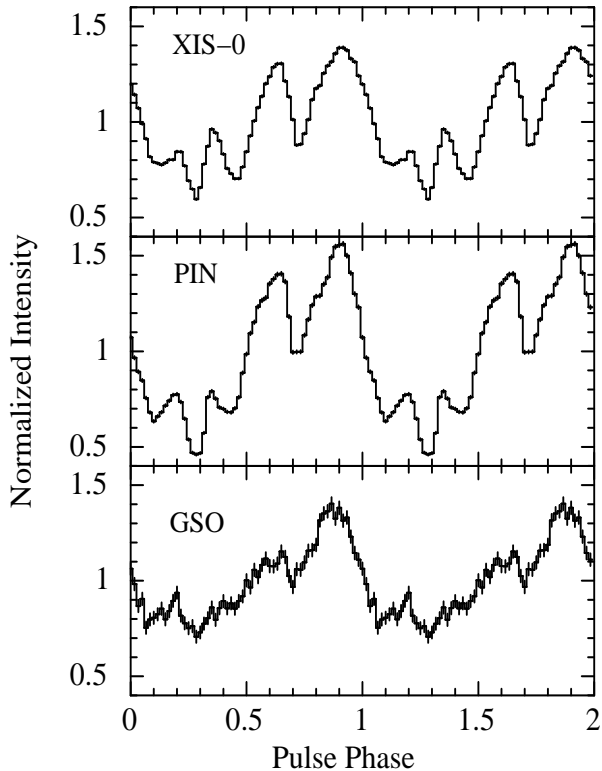}
\caption{XIS-0 (top panel; in 0.2--12 keV range), PIN (second
panel; in 10--70 keV range) and GSO (bottom panel; in 40--600 keV 
range) pulse profiles, obtained from the corresponding light curves
by using the estimated 41.4106 s pulse period of \exo. The errors in 
the figure are estimated for the 1~$\sigma$ confidence level. Two pulses
are shown for clarity.} 
\label{pp}
\end{figure}

The fifth Japanese X-ray astronomy satellite $Suzaku$ (Mitsuda et al. 2007) was 
launched on 2005 July 10. There are two sets of instruments onboard $Suzaku$ such
as X-ray Imaging Spectrometer (XIS; Koyama et al. 2007) which covers 0.2-12 keV 
energy range and Hard X-ray Detector (HXD; Takahashi et al. 2007) which covers 
10-600 keV energy range. Among the four sets of XIS, each with a 1024$\times$1024 
pixel CCD at the focus of corresponding telescope, one (XIS-1) is back-illuminated 
whereas the others are front illuminated. In full window mode, the field of view of 
the XIS is 18'$\times$18' with an effective area of 340 cm$^2$ and 390 cm$^2$ at 
1.5 keV for front-illuminated and back-illuminated CCDs, respectively. The 
non-imaging instrument HXD which was designed to detect high energy X-ray photons, 
consists of 16 identical units made up of two types of detectors such as silicon 
PIN diodes covering 10--70 keV energy range and GSO crystal scintillator covering 
30--600 keV energy range. The effective area of the PIN diodes is $\sim$ 145 cm$^2$ 
at 15 keV whereas that of GSO detectors is $\sim$315 cm$^2$ at 100 keV. As XIS-2 was 
unoperational during the observation of \exo, data from other 3 XISs are used in the 
present analysis. 

\section{Analysis and Results}

For XIS and HXD data reduction, we reprocessed the unfiltered event data
using 'aepipeline' package of HEASoft version 6.12 and utilizing the calibration 
database (CALDB) released on 2012 February 10 (for XIS) and 2011 September 13 
(for HXD). Source light curves and spectra were extracted from the reprocessed
cleaned event data of XIS, PIN and GSO detectors. The simulated background events, 
as suggested by the instrument team, were used to estimate the PIN and GSO backgrounds 
for the \exo~ observation. The response file which was released in 2008 January was 
used for HXD/PIN spectrum whereas response and effective area files released in 2010 
May were used for HXD/GSO. The reprocessed XIS event data were checked for the possible 
presence of photon pile-up. Pile-up estimation was performed by examining the Point Spread 
Function (PSF) of the three XISs by checking the count rate per one CCD exposure 
at the image peak as given by Yamada \& Takahashi\footnote{http://www-utheal.phys.s.u-tokyo.ac.jp/$~$yuasa/wiki/index.php/How to check pile up of Suzaku XIS data}. It was 
found that the XIS event data was not affected by photon pile-up. The source spectra 
were accumulated from the XIS reprocessed cleaned event data by selecting a circular 
region of $3'$ around the image centroid. The XIS background spectra were accumulated 
from the same observation by selecting circular regions away from the source. The 
response files and effective area files for XIS were generated by using the ''xissimarfgen'' 
and ''xisrmfgen'' tasks of FTOOLS. 

\begin{figure}
\centering
\includegraphics[height=2.9in, width=4.in, angle=-90]{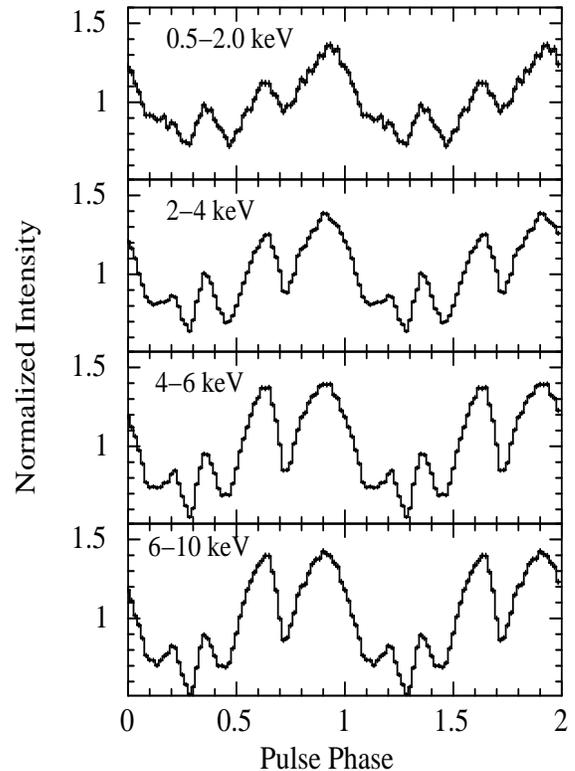}
\caption{The XIS-0 pulse profiles of \exo~ at different energy 
ranges. The presence of several dip like features in 0.3-0.7
and 0.9-1.1 pulse phase ranges are clearly seen. The error bars 
represent 1~$\sigma$ uncertainties. Two pulses are shown for clarity. } 
\label{pp1}
\end{figure}

\subsection{Timing Analysis}

\begin{figure*}
\vskip 14.2 cm
\includegraphics{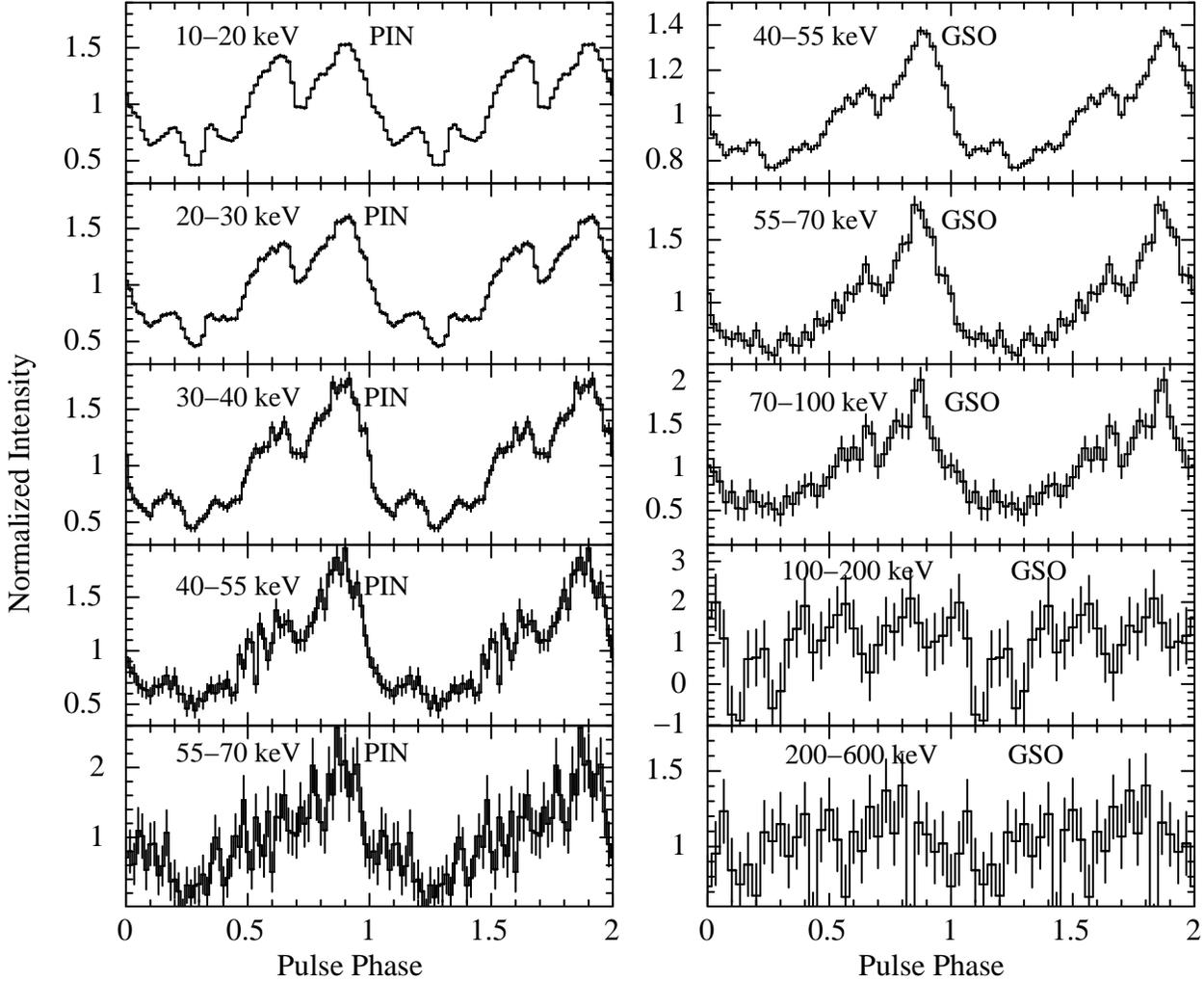}
\caption{The energy resolved pulse profiles of \exo~ obtained from
HXD/PIN and HXD/GSO light curves at various energy bands. The presence/absence of
dip like structures can be seen in 0.3-0.7 and 0.9-1.1 pulse phase ranges. The 
error bars represent 1~$\sigma$ uncertainties. Two pulses in each panel are shown 
for clarity.}
\label{pp2}
\end{figure*}

For the timing analysis, the arrival times of the X-ray photons were converted 
to the same at the solar system barycenter by using the $Suzaku$ specific 
barycenter correction task ``aebarycen''. X-ray light curves with time resolutions 
of 2~s, 1~s and 1~s were extracted from XIS (in 0.2--12~keV energy range), PIN (in 
10--70~keV energy range), and GSO (in 40--600 keV energy range) event data, 
respectively. Pulse folding and a $\chi^2$ maximization technique was applied to 
the light curves obtained from XIS, PIN and GSO event data and the pulse period
of the pulsar was estimated to be 41.4106(1) s. The light curves were then folded
with this period to get the pulse profiles at different energy range. It is found 
that the pulse profiles of the pulsar obtained from the background subtracted 
light curves of XIS-0, XIS-1 and XIS-3 are identical, whereas it is somewhat different 
to the profiles obtained from HXD/PIN and HXD/GSO data. The pulse profiles of \exo~ 
obtained from the background subtracted XIS-0, HXD/PIN and HXD/GSO light curves of 
the $Suzaku$ observation are shown in Figure~\ref{pp}. The XIS and HXD/PIN profiles, 
though look similar, the structure and depths of the dips are found to be different. 
However, the difference is very clearly visible in the HXD/GSO profile. 
To investigate the energy dependence of the pulse profile of \exo, we generated 
source and corresponding background light curves in various energy bands from the 
XIS, PIN and GSO event data. After appropriate background subtraction, the light 
curves were folded with the estimated pulse period and the corresponding pulse 
profiles are shown in Figures~\ref{pp1} \& \ref{pp2}. From the figures, it can be 
seen that the X-ray pulsations in \exo~ are clearly seen up to $\sim$100 keV. 
The dips in the profiles are very strong and clearly distinguishable up to 
$\sim$40 keV. Beyond this energy, the width and depth of the dips in the pulse
profiles decrease gradually up to $\sim$70 keV. Beyond $\sim$70 keV, however,
the dips become indistinguishable and the pulse profiles appear smooth and 
single peaked. The presence of dips in the pulse profiles up to $\sim$70 keV 
is rarely seen in the accretion powered X-ray pulsars. The characteristic 
properties of these dips in the pulse profile of \exo~ is being investigated 
in the present work through pulse phase resolved spectroscopy.

\subsection{Spectral Analysis}
\subsubsection{Pulse phase averaged spectroscopy}

We analyzed the pulse-phase-averaged spectra of \exo~ using XIS, HXD/PIN and
HXD/GSO event data. The spectra from both the front-illuminated CCDs (XIS-0 and 
XIS-3) were added together along with corresponding response matrices and 
background spectra by using package ``addascaspec''. Data from XIS-1 was 
used separately in the spectral fitting. In the spectral fitting, we selected 
data in the energy ranges of 1-10 keV for both front and back-illuminated 
CCDs (added spectra from XIS-0 and XIS-3, and XIS-1), 12-70 keV for the 
HXD/PIN and 40-200 keV for the HXD/GSO. After appropriate background 
subtraction, simultaneous spectral fitting was carried out using the 
XIS, PIN and GSO spectra with XSPEC v12.7.1. All the spectral parameters other 
than the relative instrument normalization, were tied together for all the 
detectors. The XIS spectra were binned by a factor of 5 from 1-10 keV whereas 
the PIN spectrum was not binned up to 50 keV beyond which it was binned by
a factor of 3. The GSO spectrum was binned with the fixed grouping scheme
provided by the instrument team\footnote{http://heasarc.gsfc.nasa.gov/docs/suzaku/analysis/gsobgd64bins.dat}. Because an artificial structure is known to be present in the XIS 
spectra at around the Si edge, we ignored data between 1.7--1.9~keV and 2.2--2.4 keV 
in our spectral analysis. In the beginning, we tried to fit the broad-band spectra 
of the pulsar with various continuum models such as power-law model modified with 
an exponential cut-off, Negative and Positive power law with EXponential (NPEX) 
continuum model and partial covering power-law with high energy cut-off model along 
with interstellar absorption and Iron K$_\alpha$ emission line at 6.4 keV. It is 
found that in case of all continuum models other than the partial covering high 
energy cut-off model, the spectral fitting was extremely poor with reduced $\chi^2$ 
of more than 3. Therefore, we ignored all other continuum models in our fitting.
Though a few of these models were used to describe the source
continuum earlier (for example - Reynolds et al. 1993; Wilson et al. 2008), higher 
sensitivity detectors with better energy resolution onboard $Suzaku$ and its 
broad-band spectral capability provide good statistical quality in the spectrum 
to rule out the other spectral models for this source. It may happen that the 
properties of the pulsar may be different during the normal outburst (present one) 
and giant outbursts (Reynolds et al. 1993; Wilson et al. 2008).

\subsubsection{Partial covering high energy cutoff power-law model}

The partial covering power-law with high energy cut-off model consists of two 
power-law continua with a common photon index but with different absorbing 
hydrogen column densities. The analytical form of the partially covering 
power-law with high energy cut-off model is

\begin{eqnarray}
\nonumber 
N(E) & = & {e^{-\sigma(E)N_{\mathrm H1}}}(S_{1}+S_{2}e^{-\sigma(E)N_{\mathrm H2}})
{E^{-\Gamma}}{I(E)}
\end{eqnarray}

where
\begin{eqnarray}
\nonumber  I(E) & = & 1  \hspace{0.83in} for ~E < E_\mathrm c \\
\nonumber       & = & e^{- \left({E-E_\mathrm c}\over{E_\mathrm f}\right)} \hspace{0.27in}  for ~E > E_\mathrm c
\end{eqnarray}

$N_{H1}$ is the Galactic equivalent hydrogen column density, $N_{\mathrm H2}$ is the 
additional equivalent hydrogen column density of the material local to the neutron star, 
$N(E)$ is the observed intensity, $\Gamma$ is the photon index, $\sigma$ is the photo-electric 
cross-section, $S_{1}$ and $S_{2}$ are the normalizations of the power-law, 
$E_{\mathrm c}$ is the cut-off energy and $E_{\mathrm f}$ the e-folding energy. The 
covering fraction of the absorbed power-law due to the presence of additional matter local
to the neutron star is expressed as Norm$_2$ / (Norm$_1$ + Norm$_2$) = $S_2 / (S_1 + S_2)$].

\begin{figure}
\centering
\includegraphics[height=2.85in, width=2.2in, angle=-90]{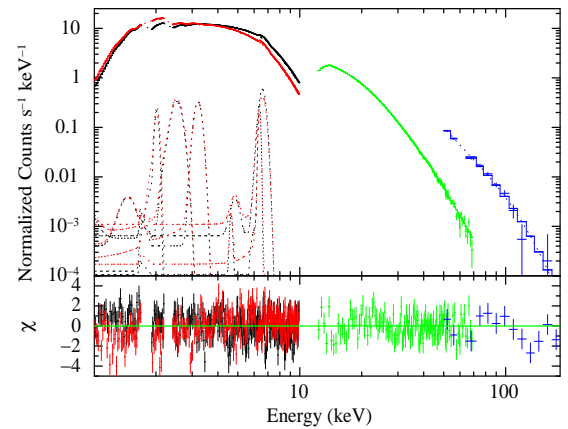}
\caption{Energy spectra of \exo~ obtained with the XIS, PIN and GSO 
detectors of the $Suzaku$ observation at the peak of the type~I outburst, 
along with the best-fit model consisting of a partially absorbed power-law
with high energy cutoff continuum model with neutral absorber. Five emission 
lines (as given in Table~1) are detected in the broad-band spectrum of the 
pulsar. The contributions of the residuals to the $\chi^2$ for each energy 
bin for the best-fit model are shown in the bottom panel.}
\label{spec-fg1}
\end{figure}

\begin{figure}
\centering
\includegraphics[height=2.85in, width=2.2in, angle=-90]{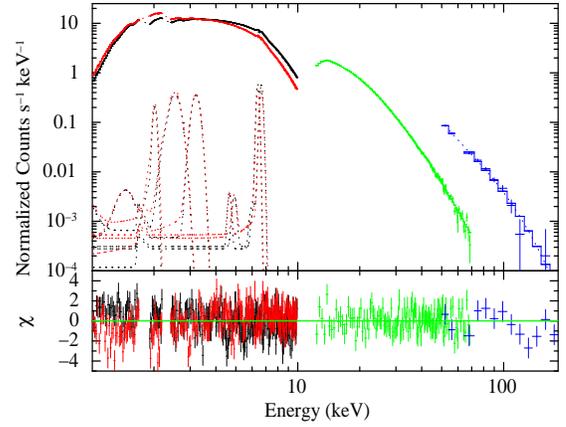}
\caption{Energy spectrum of \exo~ obtained with the XIS, PIN and GSO 
detectors of the $Suzaku$ observation at the peak of the type~I outburst, 
along with the best-fit model comprising a partially absorbed power-law with
high energy cutoff continuum model with partially ionized absorber. The emission 
lines (as given in Table~1) detected in the pulsar spectrum are represented with 
Gaussian functions in the figure. The contributions of the residuals to the 
$\chi^2$ for each energy bin for the best-fit model are shown in the bottom panel.}
\label{spec-fg2}
\end{figure}

As mentioned earlier, the relative instrument normalizations of added front-illuminated
CCDs (XIS-0 and XIS-3, quoted as XIS03), XIS-1, PIN and GSO detectors were kept free. 
The corresponding values obtained are 1.00:1.04:0.99:1.02 for XIS03:XIS1:PIN:GSO
with a clear agreement with the values at the time of detector calibration. After fitting
the spectra with above model, excess residuals were found to be present at $\sim$2 keV, 
$\sim$2.5 keV, $\sim$3.2 keV and $\sim$6.6 keV. As the observation was during the peak 
of the Type~I outburst and the pulsar being bright at hard X-rays, the presence of several 
emission lines is expected in the spectrum. Therefore, we added Gaussian functions at above 
energies to the spectral model and re-fitted the spectra. Though these emission lines are 
weak, addition of these lines to the model improved the simultaneous spectral fitting with 
reduced $\chi^2$ of 1.59 (for 638 dof). Detection of several emission lines at soft X-rays 
and the pulsar being so bright during the $Suzaku$ observation, we attempted to fit the 
broad-band spectra with the partial covering high energy cut-off power-law model with 
partially ionized absorber ($zxipcf$ model in XSPEC; Reeves et al. 2008) and 
Gaussian functions for detected emission lines. Using the partially ionized 
abosrber component in the spectral model, the ionization states of the absorbing 
medium and the corresponding covering fraction could be investigated. The $zxipcf$
model component in $XSPEC$ uses a grid of $XSTAR$ (Kallman et al. 2004) photoionized 
absorption models for absorption. The free parameters in this model are $N_{H}$ (in 
10$^{22}$ cm$^{-2}$), $C$ (covering fraction), and the ionization parameter $\xi$ (erg 
cm s$^{-1}$; Reeves et al. 2008). The spectral fitting was improved marginally with 
reduced $\chi^2$ of 1.51 (for 637 dof). The energy spectra of the pulsar \exo~ are 
shown in Figure~\ref{spec-fg1} \& \ref{spec-fg2} (for partial covering models with 
neutral and partially ionized absorbers, respectively) along with the spectral 
components (top panels) and residuals to the best-fit model (bottom panels). 
As the difference in both the spectral models is the nature of the additional absorption 
components without any noticable change in the best-fit value of other spectral parameters, 
both the figures look similar. The best-fit parameters obtained from the simultaneous 
spectral fitting to the XIS, PIN and GSO data with partial covering model modified with 
neutral and partially ionized absorbers are given in Table~\ref{spec_par}. 

The spectral fitting of the $RXTE$ observations of the pulsar during 2006 June giant 
outburst showed the presence of a cyclotron resonance feature centered at $\sim$11 
keV (Wilson et al. 2008). A cyclotron resonance line at $\sim$63 keV was reported at
certain pulse phase ranges of the pulsar from the $INTEGRAL$ observations during same 
giant outburst (Klochkov et al. 2008). From $RXTE$ observations of the pulsar at 
relatively lower luminosity level, a spectral feature was detected at $\sim$36 keV 
and ascribed to a cyclotron absorption feature (Reig \& Coe 1999). However, 
in our spectral fitting, no such absorption feature was present at above energies.
Therefore, we did not include any additional cyclotron absorption component to the 
spectral fitting.

\begin{table}
\centering
\caption{Best-fit parameters of the phase-averaged spectra for \exo~
during 2007 May $Suzaku$ observation with 1$\sigma$ errors. Model-1 : Partial covering 
(with neutral absorber) high energy cutoff power-law model with five Gaussian components, 
Model-2 : Partial covering (with partially ionized absorber) high energy cutoff power-law 
model with five Gaussian components.}
\begin{tabular}{llll}
\hline
Parameter      		&\multicolumn{2}{|c|}{Value} 	 \\
				 &Model-1 	&Model-2 \\
\hline
N$_{H1}$  			     &2.04$\pm$0.01		&2.07$\pm$0.01  \\
N$_{H2}$ 			     &40.9$\pm$3.3		&45.4$\pm$2.8	\\
Covering Fraction                    &0.13$\pm$0.01		&0.23$\pm$0.01	\\
$log(\xi)$			     &------			&1.56$\pm$0.15  \\
Power-law index 		     &1.33$\pm$0.01		&1.37$\pm$0.01  \\
High energy cutoff (keV)	     &12.4$\pm$0.2		&11.0$\pm$0.8   \\
E-fold energy (keV)		     &21.2$\pm$0.2		&21.5$\pm$0.2   \\
\\
Emission lines			     &                          &\\
\\
Si~XIII\\
Line energy (keV)		     &2.01$\pm$0.01		&2.01$\pm$0.01\\
Line width (keV)		     &0.001			&0.001\\
Line eq. width (eV)		     &2				&2$\pm$2\\
\\
Si~XIV\\
Line energy (keV)		     &2.50$\pm$0.03		&2.51$\pm$0.02\\
Line width (keV)		     &0.13$\pm$0.03		&0.14$\pm$0.02\\
Line eq. width (eV)		     &11			&13\\
\\
S~XV\\
Line energy (keV)		     &3.19$\pm$0.01		&3.19$\pm$0.01\\
Line width (keV)		     &0.11$\pm$0.02		&0.12$\pm$0.01\\
Line eq. width (eV)		     &8				&10\\
\\
Iron K$_\alpha$\\
Line energy (keV)		     &6.39$\pm$0.01		&6.4$\pm$0.02\\
Line width (keV)		     &0.01$^{+0.02}_{-0.01}$	&0.05$\pm$0.02\\
Line eq. width (eV)		     &7				&19\\
\\
Fe~XXVI\\
Line energy (keV)		     &6.61$\pm$0.01		&6.66$\pm$0.02\\
Line width (keV)		     &0.20$\pm$0.02		&0.08$\pm$0.03\\
Line eq. width (eV)		     &51			&25\\
\\
Source flux \\
1 - 10 keV range		     &2.94$^{+0.07}_{-0.06}$	&2.94$^{+0.08}_{-0.09}$\\
10 - 70 keV range		     &5.92$^{+0.13}_{-0.12}$	&5.92$^{+0.15}_{-0.13}$\\
70 - 200 keV range		     &2.53$^{+0.05}_{-0.05}$    &2.54$^{+0.04}_{-0.07}$\\
\\
Reduced $\chi^2$                     &1.59 (638)		&1.51 (637)  \\
\hline
\end{tabular}
\\
N$_{H1}$ = Galactic equivalent hydrogen column density along the line of sight, N$_{H2}$ = Additional hydrogen column density of material local to the neutron star. 
N$_{H1}$ and N$_{H2}$ are in the units of (10$^{22}$ atoms cm$^{-2}$). Source flux is in the unit of 10$^{-9}$ ergs cm$^{-2}$ s$^{-1}$. Source flux quoted above is not corrected for interstellar absorption.
\label{spec_par}
\end{table}

\subsubsection{Pulse phase resolved spectroscopy}

\begin{figure*}
\vskip 14.2 cm
\includegraphics{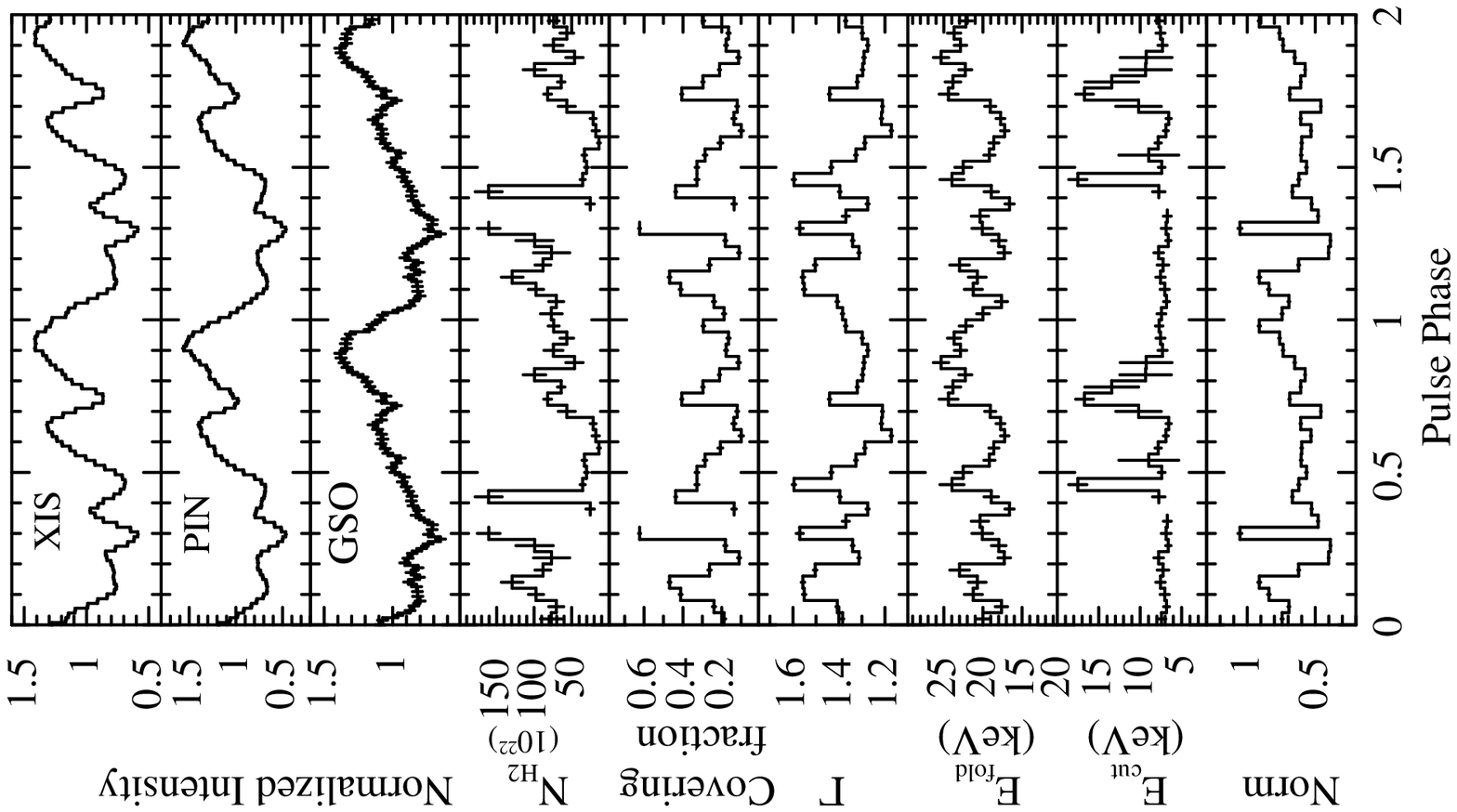}
\includegraphics{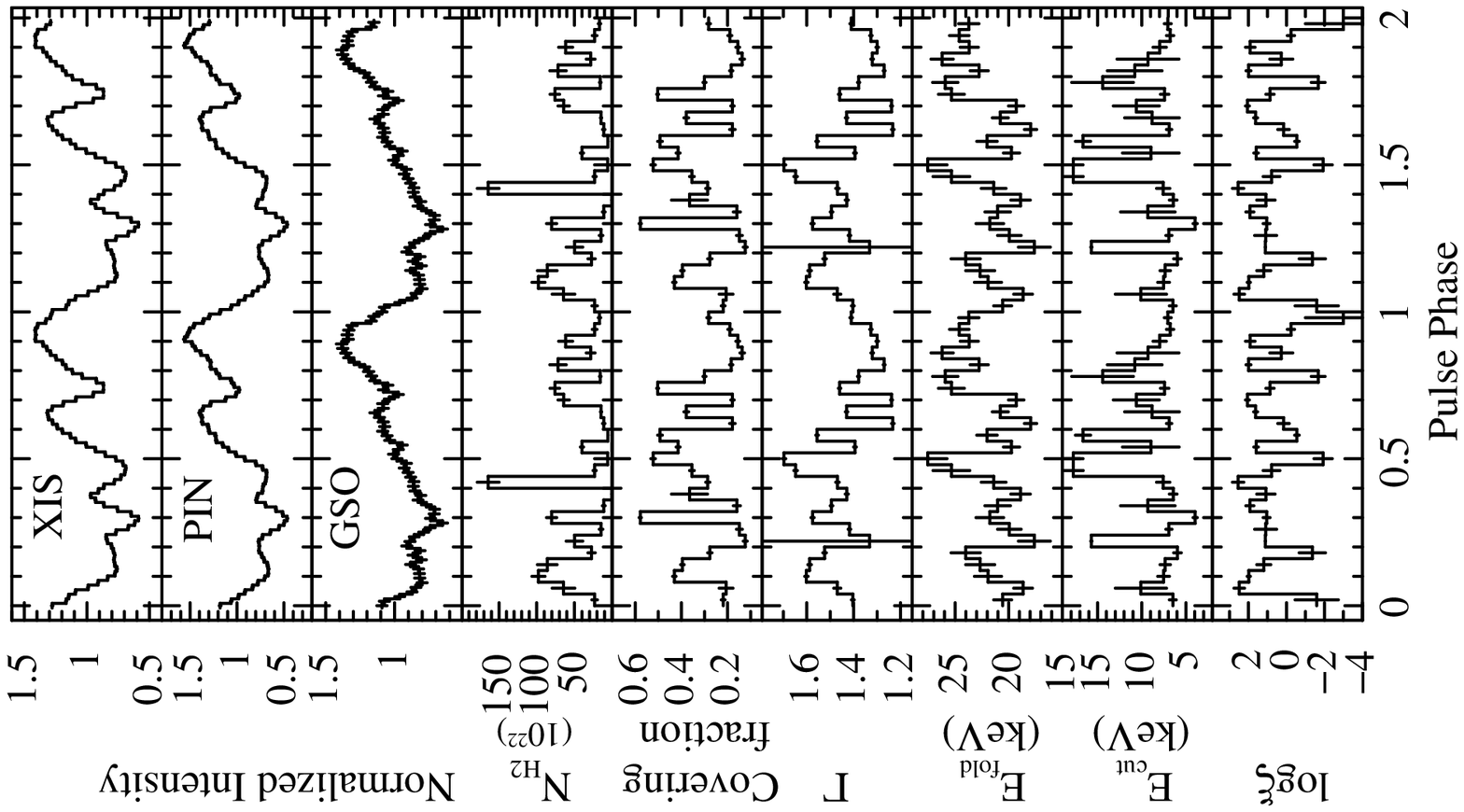}
\caption{Best-fit spectral parameters obtained from the phase resolved
spectroscopy of $Suzaku$ observation of \exo. The XIS (in 0.2-12 keV energy 
range), PIN (in 10-70 keV energy range) and GSO (in 40-600 keV energy range) 
pulse profiles of the pulsar are shown in top three panels of both sides of the 
figure, respectively. The other panels in the left side show the spectral
parameters obtained from pulse phase-resolved spectroscopy by using partial 
covering model with neutral absorber whereas the panels in the right side 
shows the same for partial covering model with partially ionized absorber. 
The errors shown are estimated for 1$\sigma$ confidence 
level. }
\label{phrs}
\end{figure*}

The presence of several prominent dips in the pulse profile and partial covering
power-law with high energy cut-off being the best fit continuum model imply 
the possibilities of presence of streams of matter at various pulse phases 
around the poles of the transient Be/X-ray binary pulsar \exo. To investigate this, 
pulse-phase-resolved spectroscopy was performed on the XIS and PIN data. 
By applying phase filtering in the FTOOLS task ``XSELECT'', we accumulated
25 pulse phase resolved spectra from XIS and HXD/PIN event data. Data from 
GSO detectors were not used in phase-resolved spectroscopy as the 
signal-to-noise ratio was very poor for spectra of each phase bins. 
For phase-resolved spectroscopy, we used same background spectra and 
response matrices for corresponding detectors as were used for phase-averaged 
spectroscopy. Simultaneous spectral fitting was carried out in the 1.0-70.0 keV 
energy band to each of the 25 phase resolved XIS and PIN spectra. The phase 
resolved XIS and PIN spectra were fitted with the partial covering power-law with 
high energy cut-off model with neutral and partially ionized absorber (as was done 
in phase-averaged spectroscopy). Some of the parameters such as the values of 
relative instrument normalizations, the Galactic absorption ($N_{H1}$), center 
energy and width of emission lines were fixed at the values obtained from 
phase-averaged spectroscopy. The parameters obtained from the simultaneous 
spectral fitting to the XIS and PIN phase resolved spectra using partial 
covering model with neutral and partially ionized absorber are shown in the 
left and right panels of Figure~\ref{phrs}, respectively. The top three panels 
in both the sides of Figure~\ref{phrs} show the pulse profiles of \exo~ obtained 
from XIS, PIN and GSO data. The values of additional column density (N$_{H2}$) 
and covering fraction over pulse phases are shown in fourth and fifth panels 
of the figure, respectively. The change in values of power-law photon index, 
e-folding energy and cut-off energy over pulse phases are shown in sixth, 
seventh and eighth panels respectively. The left side bottom panel shows 
the variation of power-law normalization whereas the right side bottom panel 
of the figure shows the value of ionization parameter $log~\xi$ over pulse 
phases (for the partial covering model with partially ionized absorber). 

Among the spectral parameters, most notable and systematic variability is 
seen in the values of additional column density and covering fraction of the 
absorber. The values of N$_{H2}$ are found to be high at phases where dips 
are present in the pulse profiles. This pattern is seen in case of both 
the models i.e. partial covering model with neutral as well as partially 
ionized absorber. In case of partial covering model with partially ionized 
absorber, the value of the ionization parameter was found to be relatively 
higher at dip phases indicating the presence of highly ionized additional 
absorber at several pulse phases causing dips in the pulse profiles. The 
higher values of cut-off energy were also found at pulse phases that are 
coincident with the dips in the pulse profiles. The values of the power-law 
photon index and e-folding energy are found to be variable over the pulse 
phases of the pulsar. However, it is difficult to correlate these 
variabilities with the dips in the pulse profiles. We did not find any 
variability in the flux of both the iron emission lines over the pulse 
phases of the pulsar. This suggests that the matter emitting the iron 
fluorescence lines is probably distributed symmetrically around the pulsar. 

\section{Discussion}

Be/X-ray binaries usually show two types of X-ray outbursts such as normal (type~I) 
outbursts and giant (type~II) outbursts. The normal outbursts are characterized by 
lower luminosity ($\sim$10$^{36-37}$ erg s$^{-1}$) and occur near the periastron 
passage. However, the giant outbursts are characterized by high luminosities 
($\geq$10$^{37}$ erg s$^{-1}$) and very rare (Stella, White \& Rosner 1986; 
Negueruela et al. 1998; Bildsten et al. 1997). In case of Be/X-ray pulsar \exo, 
the type~I outbursts are seen almost every $\sim$46 days (orbital period) 
whereas the giant outbursts were only seen twice; in 1985 during which the 
pulsar was discovered (Parmar et al. 1989a, 1989b) and in 2006 (Wilson 
et al. 2008). The pulsar has been monitored extensively with various X-ray 
observatories giving rise to the accurate estimation of the orbital parameters 
of the binary system (Wilson et al. 2008). Attempts have been made to measure 
and understand the complex nature of the phase-averaged pulsar spectrum (Klochkov 
et al. 2008; Wilson et al. 2008 and references therein). Phase-resolved spectroscopy 
has been performed only once when the pulsar was undergoing the second giant outburst 
in 2006 (Klochkov et al. 2008). Though, phase resolved spectral analysis was carried 
out on the $RXTE$ observations of the pulsar during a normal outburst (Reig \& Coe 1998), 
the inferior spectral capability and low energy threshold of the detectors limited the 
understanding of the complex nature of the pulse profile. Broad-band capability and high 
sensitivity of detectors onboard $Suzaku$ provided the opportunity to investigate the 
properties of the pulsar at different pulse phases in more detail. 

\subsection{Pulse Profile}

The pulse profile of the transient pulsar \exo~ are found to be strongly energy 
dependent. The presence of several prominent dips makes the soft X-ray pulse profiles 
complex. The strength of these dips gradually decreases with energy, making the 
profile a smooth and single peaked at high energy bands. X-ray pulsations in \exo~ are 
detected up to as high as $\sim$100 keV. The shape of the pulse profiles obtained 
from $Suzaku$ observation are found to be different from that obtained from $RXTE$ 
(Reig \& Coe 1998; Sasaki et al. 2010), $JEM-X$ (Martinez Nunez et al. 2003), $INTEGRAL$ 
(Klochkov et al. 2008) observatories, though a few of these observations were carried 
out during the 2006 giant outburst. The presence of dips in the pulse profile are seen 
in other transient X-ray pulsars such as A0535+262 (Naik et al. 2008), GX~304-1 (Devasia 
et al. 2011), GRO~J1008-57 (Naik et al. 2011), 1A~1118-61 (Devasia et al. 2011; Maitra et 
al. 2012), RX~J0440.9+4431 (Usui et al. 2012), Vela~X-1 (Maitra \& Paul 2012) etc. In most 
of the cases, pulse-phase resolved spectral analysis showed the presence of additional
absorption component at certain pulse phases that partially obscured the emitted radiation
giving rise to dips in the pulse profiles. The additional absorption is understood to be 
taking place by matter in the accretion streams that are phase locked with the neutron star.

In accretion powered pulsars, pulse profiles at hard X-rays are generally simpler
and smoother than that at low energies. Low-energy pulse profiles are more affected
by absorption by circumstellar matter and/or additional matter distribution near
the neutron star. In case of majority of the Be/X-ray binary pulsars, the soft X-ray 
pulse profiles are found to consist of dips at various pulse phases whereas the hard 
X-ray pulse profiles are smooth and single-peaked. \exo~ is one of a few other accreting 
X-ray pulsar which clearly show hard X-ray pulsations. In the present work, it was found 
that the 41.41 s pulsations are detected in the X-ray light curves at high energies 
($\sim$100 keV). Similar results are seen in a few other pulsar such as
1A~1118-61 (up to $\sim$100 keV; Coe et al. 1994; Bildsten et al. 1997), 
2S~1417-624 (up to $\sim$100 keV; Finger et al. 1996), GX~1+4 (up to $\geq$100 
keV; Naik, Paul \& Callanan 2005), GRO~J1008-57 (up to $\sim$100 keV; Naik et al. 2011) 
etc. High energy photons being less affected by the absorption/scattering by matter in the
interstellar medium as well as matter distribution close to the neutron star, the hard 
X-ray pulse profiles possibly indicate the intrinsic radiation pattern from the pulsar.
Analyzing the shape of energy resolved pulse profiles of \exo~ during its second giant
outburst in 2006, Sasaki et al. (2010) modeled the geometry of the neutron
star by identifying the emission components of the magnetic poles. The asymmetric shape 
of the pulse profiles during the giant outburst was explained in terms of moderately 
distorted magnetic field the consequence of which is that one pole of the pulsar gets 
closer to the line of sight than the other. Because of the asymmetry in position of the
poles, the symmetric pulse profiles from both the poles merged together and appeared 
as asymmetric in shape. Similar analysis was carried out to understand the geometry of
the neutron star in other binary pulsars such as 4U~0115+63 and V0332+53 at various 
luminosity levels of their outbursts (Sasaki et al. 2012). In contrast to the results 
obtained from the 2006 giant outburst of \exo~ (Klochkov et al. 2008; Sasaki et al. 
2010), we find that at high energies, the shape of the pulse profiles appeared to 
be symmetrical. At soft X-rays, the pulse profile was found to be complex because 
of the presence of several absorption dips. The shape of the observed pulse profiles 
during type~I outburst suggests the asymmetric distribution of the magnetic poles 
which was used to describe the findings during the giant outburst in EXO~2030+375
may not be applicable.

\subsection{Spectroscopy}

The broad-band X-ray spectrum of \exo~ in 1-200 keV energy range has been described 
here in this paper for the first time in detail. Camero Arranz et al. (2005) presented the 
pulsar spectrum in 5-300 keV energy range obtained from the $INTEGRAL$ observation
of a type~I outburst in 2002 December by a model consisting of a disk blackbody ($kT$ 
$\sim$ 8 keV) and a power-law with $\Gamma$ of 2.04 or a Comptonized component. The lower
threshold at 5 keV limited the understanding of the characteristic properties of
the pulsar at soft X-rays. A blackbody component at a temperature of 8 keV is rarely
reported so far and difficult to explain in accretion powered X-ray pulsars. They also 
did not find any evidence of presence of an iron line or cyclotron line features in the 
pulsar spectrum. $RXTE$ observations of the pulsar during type~I outbursts, though used
a blackbody and power-law components to explain the 2.7-30 keV spectrum, the blackbody
temperature was estimated to be $\leq$1.4 during entire outburst (Reig \& Coe 1999).
The presence of an iron emission line at 6.4 keV was seen in the $RXTE$ spectra. Apart
from the $RXTE$ observations, there are other observations at high energy bands some of
which are during the two giant outbursts. 

In case of Be/X-ray binaries, the circumstellar
disk around the Be companion plays an important role in the X-ray emission from the 
neutron star. During the periastron passage, the outer edge of the disk is truncated 
by the neutron star resulting in X-ray outbursts. During the periastron passage, as 
additional matter is being accreted onto the neutron star, the value of the equivalent
hydrogen column density increases significantly. This increase in column density modifies the 
emitted radiation at soft X-rays. While simultaneous spectral fitting to the pulsar
spectrum in 1-200 keV energy range, the estimated value of column density always exceeded
the corresponding Galactic value in the direction of the pulsar. Even though, we attempted
to fit the broad-band spectra of \exo~ with various continuum models, partial covering
model provided best-fit to the data. Though the estimated value of absorption column density
was higher, another significant absorption component with certain value of covering fraction
(as given in Table~1) was required in the spectral fitting. \exo~ being a bright hard X-ray
pulsar and presence of significant amount of matter at periastron passage, there is a 
possibility of ionizing the surrounding matter. To investigate this, we attempted to fit
the spectrum with an partially ionized absorber which yielded marginal improvement in the 
spectral fitting. Apart from this, low energy emission lines are also detected in the pulsar
spectrum. We found Si~XIII, Si~XIV, S~XV, Fe K$_\alpha$ and Fe~XXVI emission lines in
the XIS spectra of the pulsar. Though the multiple iron emission lines are seen in other 
accretion powered X-ray pulsars e.g. Cen~X-3 (Naik, Paul, \& Ali 2011 and references 
therein), GX~1+4 (Naik, Paul \& Callanan 2005 and references therein), the low energy 
Si and S emission lines are detected for the first time in this pulsar. As the previous 
observations were carried out during giant outbursts or with the instruments of lower 
spectral capability at soft X-rays, these Si and S lines were missed out from detection. 
During giant outbursts which are not associated with the periastron passage, lack of 
significant amount of additional matter in the close proximity of the neutron star as 
detected during type~I outbursts, reduces the chance of presence/detection of these 
emission lines in the spectrum. The spectral fitting of $Suzaku$ observation of the 
pulsar did not show any evidence of presence of cyclotron absorption features at 
earlier reported energies from other observations.

Pulse-phase resolved spectroscopy of the $Suzaku$ observation of \exo~ showed that the value
of additional column density ($N_{H2}$) due to the material local to the neutron star
was about two orders of magnitude higher at certain pulse phases. It can be seen (from
the top panels of Figure~\ref{phrs}) that absorption dips are present at same pulse phase
ranges. The additional high value of $N_{H2}$, therefore, explains the presence of 
absorption dips in the pulse profile. Apart from the variation of the additional column 
density and covering fraction over pulse phases, the value of high energy cut-off was found 
to be higher at dip phases. However, other parameters did not show any systematic variation 
over the pulse phases.

\section{Conclusion}
In this paper, we performed timing and broad-band spectral analysis on the $Suzaku$ 
observation of the Be/X-ray transient pulsar \exo~ during the peak of a type~I outburst. 
The  41.41 s pulsations were detected in the light curves up to as high as $\sim$100 keV.
The pulse profiles are found to be strongly energy dependent. Narrow dips which are 
generally seen in the pulse profiles of accretion powered pulsars up to $\sim$10 keV, are 
seen in the profiles up to as high as ∼70 keV. At soft X-rays, the shape of the pulse 
profiles is found to be complex whereas at high energies, it appeared symmetrical. The 
complex nature of the profiles at soft X-rays is interpreted as because of the presence 
of several absorption dips. A partial covering power-law with high energy cut-off continuum 
(with neutral as well as partially ionized absorber) model was found to be the preferred 
model to describe the broad-band spectrum in 1.0-200.0 keV energy range. High values of 
additional column density at the dip phases in the pulse profile confirm the presence of 
stream of absorbers that are phase locked with the pulsar. Apart from the fluroscence 
iron emission line, several low energy lines from S and Si are also detected in the pulsar 
spectrum. Cyclotron resonance scattering features, though reported earlier in the spectrum 
of this pulsar, are not detected in the 1.0-200.0 keV spectrum of the pulsar. Pulse-phase 
resolved spectroscopy revealed that the higher values of ionization parameter at the dip 
phases of the pulse profile may be the cause of the presence of dips up to $\sim$70 keV.

\section*{Acknowledgments}
We thank the referee for his/her suggestions that improved the presentation of the 
paper. The research work at Physical Research Laboratory is funded by the Department of
Space, Government of India. The authors would like to thank all the members of the 
$Suzaku$ for their contributions in the instrument preparation, spacecraft operation, 
software development, and in-orbit instrumental calibration. This research has made 
use of data obtained through HEASARC Online Service, provided by the NASA/GSFC, in 
support of NASA High Energy Astrophysics Programs.

\end{document}